\documentclass[prd,twocolumn,aps,showpacs,nofootinbib,nobibnotes,superscriptaddress,preprintnumbers]{revtex4-1}
\usepackage{epsfig}
\usepackage{graphicx}
\usepackage{bm}
\usepackage[usenames, dvipsnames]{color}
\usepackage{dcolumn}   
\usepackage[spanish,english]{babel}
\usepackage{bm}     
\usepackage{bbm}       
\usepackage{amssymb}  
\usepackage{amsmath}
\usepackage{latexsym}
\usepackage{float}
\usepackage{ifthen}
\usepackage{caption,subfig}
\usepackage{enumerate}
\usepackage{url}
\usepackage{caption,subfig}
\usepackage{amsopn}
\usepackage{hyperref}

\usepackage{amsfonts}
\usepackage{multirow}
\usepackage{array}
\usepackage{booktabs}
\usepackage{rotating}

\usepackage{ulem}
\def\clap#1{\hbox to 0pt{\hss#1\hss}}

\def\bea{\begin{eqnarray}}
\def\eea{\end{eqnarray}}
\def\be{\begin{equation}}
\def\ee{\end{equation}}
\def\mpl{M_{\rm P}}

\begin{document}
%\hspace{5.2in} \mbox{NORDITA-2015-38}\\\vspace{1.53cm} % Preprint number

\title{Horndeski in the Swampland}

\author{Lavinia Heisenberg} \email{lavinia.heisenberg@phys.ethz.ch}
\affiliation{Institute for Theoretical Physics, 
ETH Zurich, Wolfgang-Pauli-Strasse 27, 8093, Zurich, Switzerland}
 
\author{Matthias Bartelmann} \email{bartelmann@uni-heidelberg.de}
\affiliation{Universit\"at Heidelberg, Zentrum f\"ur Astronomie, Institut f\"ur Theoretische Astrophysik, Germany}

\author{Robert Brandenberger} \email{rhb@hep.physics.mcgill.ca}
\affiliation{Physics Department, McGill University, Montreal, QC, H3A 2
T8, Canada}

\author{Alexandre Refregier} \email{alexandre.refregier@phys.ethz.ch}
\affiliation{Institute for Particle Physics and Astrophysics, Department of Physics, ETH Zurich, Wolfgang-Pauli-Strasse 27, 8093, Zurich, Switzerland}

\date{\today}

\begin{abstract}
We investigate the implications of string Swampland criteria for alternative theories of gravity.
Even though this has not rigorously been proven, there is some evidence that exact de-Sitter solutions with 
a positive cosmological constant cannot be successfully embedded into string theory, and that the 
low energy effective field theories containing a scalar field $\pi$ with a potential $V$ in the habitable landscape should satisfy 
the Swampland criteria $|V'|/V\ge c\sim\mathcal{O}(1)$. 
As a paradigmatic class of modified gravity theories for inflation and dark energy, we consider
 the extensively studied family of Horndeski Lagrangians in view of cosmological observations.
 Apart from a possible potential term, they contain derivative self-interactions as the Galileon and 
 non-minimal couplings to the gravity sector. Hence, our conclusions on the Galileon sector 
 can be also applied to many other alternative theories with scalar helicity modes containing derivative interactions
 such as massive gravity and generalized Proca.
In the presence of such derivative terms, the dynamics of the scalar mode is substantially modified,
and imposing the cosmological evolution constrained by observations places tight constraints on $c$ within the Swampland
conjecture.

\end{abstract}

%\pacs{95.35.+d, 04.50.Kd}
%PACS NEEDED

\maketitle

%-----------------------------------------------------------------

\section{Introduction}

The standard cosmological paradigm relies on the existence of hypothetical scalar fields causing two phases of accelerated expansion during the evolution history of the universe. First of all, cosmological inflation is a period of extremely rapid expansion of the universe that is thought to have taken place immediately after the Big Bang. To explain the dynamics of inflation, a scalar quantum field is needed, which is spatially homogeneous and has a finite energy density. If the field changes sufficiently slowly in time, it has negative pressure and effectively behaves like a cosmological constant, thus causing the expansion of the universe to accelerate. Secondly, the present universe seems to have entered into a similar phase of accelerated expansion. The explanation of this observed acceleration is the subject of current research and has led to the concept of dark energy. The $\Lambda$CDM model is a cosmological model that describes the development of the universe since the Big Bang with a few parameters. It is the simplest model that is in good agreement with virtually all cosmological measurements. $\Lambda$ stands for the positive cosmological constant. Extensions of this simple model typically entail a time-evolving scalar field. Up to now, cosmological observations show no significant deviation of the cosmic acceleration from that expected from a cosmological constant, so we seem to be living in a universe that is either of de-Sitter type or close to it.

Einstein's theory of General Relativity (GR) describes the interaction between matter on the one hand and space and time on the other. It interprets gravity as the geometric property of the curved four-dimensional space-time. GR extends special relativity and Newtonian gravity and has been experimentally confirmed in numerous tests. However, in order to account for inflation and time evolving dark energy, one has to introduce additional fields beyond the standard model and extend the underlying theory accordingly. In this context, modifications of gravity have been considered, with the most studied ones containing an additional scalar field. As a paradigmatic class of modified gravity theories for inflation and dark energy, one can study scalar-tensor theories and, in particular, the extensively studied family of Horndeski Lagrangians \cite{Horndeski:1974wa}. One characteristic property of these theories is the second-order nature of the derivative self-interactions of the scalar field that, in turn, require the presence of derivative non-minimal couplings to the gravity sector (see e.g.\ \cite{Heisenberg:2018vsk} for a recent review). Most of our effective field theories of gravity, GR or beyond, face the same tenacious challenge concerning their consistent UV completion into a quantum gravity theory.
A promising candidate for quantum gravity is string theory in view of its successful unification of the standard model of particle physics with gravity. Should string theory be the ultimate theory of quantum gravity, one pertinent question would be whether our constructed effective field theories of gravity can naturally be embedded into string theory. 

One can divide the effective field theories into two groups: the \emph{Landscape}, where field theories can successfully be embedded into string theory, and the \emph{Swampland} as an inhabitable region, where field theories are incompatible with quantum gravity \cite{OoguriVafa}. Motivated by string-theoretical constructions, a new de-Sitter Swampland conjecture was postulated recently \cite{Obied:2018sgi}, asserting that any scalar field arising from string theory should satisfy a universal bound on its potential $|V'|\ge \frac{c}{\mpl}V$. One immediate consequence of this conjecture is that (meta-)stable de-Sitter vacua in string theory would be excluded. The existence of stable de-Sitter vacua in critical string theory has been questioned in the literature before \cite{noLambda}, however metastable de-Sitter vacua might be possible. The cosmological implications of this new \emph{Swampland conjecture} are multifaceted, in particular, Quintessence-type models for inflation and dark energy are highly constrained \cite{CosmoImpl} (see also \cite{recent} for some other related discussions). 

These string-theory criteria can also be used to constrain alternative theories of gravity. Still remaining within the class of theories containing one additional scalar field, we consider the Horndeski scalar-tensor theories with derivative self-interactions. They can be applied to both inflationary scenarios and concrete dark-energy models. The Quintessence models are very limited theories with just a potential term of the scalar field, whereas the Horndeski models represent the most general Lagrangians for a scalar field in the presence of derivative interactions. They contain the Galileon interactions as a subclass \cite{Nicolis:2008in}, which is part of many modified gravity theories including massive gravity \cite{deRham:2010kj} and generalized Proca theories \cite{HeisenbergProca}. Thus, even though we specifically study the implications of the string Swampland criteria for the Horndeski scalar field, these implications will also be directly applicable to the longitudinal mode of many other modified gravity theories, at least concerning their derivative interactions. In fact, the decoupling limit of massive gravity can be covariantized and the resulting theory belongs to a subclass of Horndeski theories \cite{deRham:2011by}. Of course both massive gravity and generalized Proca theories contain important non-trivial interactions that go beyond the Galileon interactions and the associated degrees of freedom descend from a full-fledged tensor and vector field respectively. The presence of the additional helicity modes will have important implications for the Swampland conjectures, that go beyond the scope of our present work.

It is well known that the Galileon interactions are protected from quantum corrections due to their antisymmetric structure. In order for them to have a local, analytic Wilsonian UV completion, the positivity requirements of the tree level scattering amplitudes impose for instance a constraint between the quartic and cubic interactions and in \cite{deRham:2017imi} it has been shown that there is no obstruction to a local UV completion. Going beyond the Galileon interactions, the class of Horndeski theories where the Galileon invariance is weakly broken, is insensitive to loop corrections on quasi de Sitter backgrounds \cite{Pirtskhalava:2015nla}.

\section{Horndeski and the String Swampland}

\begin{figure}[t]
  \includegraphics[height=3.0in,width=3.0in]{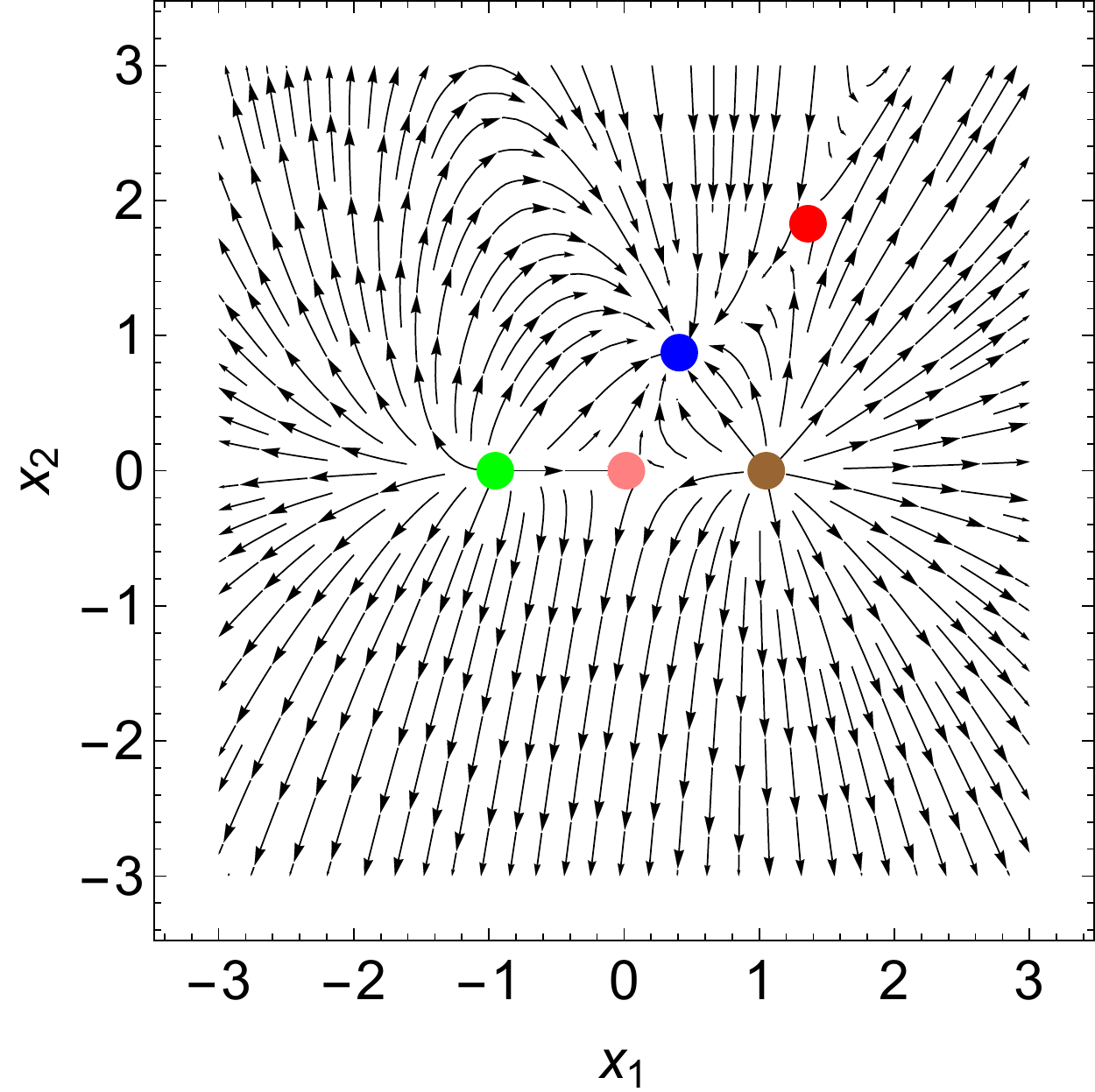}
\caption{The figure shows the phase map of the autonomous system for the model with $\alpha_3=0$ and $\alpha_4=0$. The trajectories of $\{ x_1,x_2\}$ on the $y=0$ plane are shown for $\alpha=-0.02$ and $\lambda=0.9$. The critical points $II$ (green), $III$ (brown), $IV$ (pink), $VI$ (red) and $VII$ (blue) are highlighted. The latter represents the scalar-field-dominated epoch, whereas $IV$ represents the matter-dominated phase. The critical point $I$ (corresponding to the radiation-dominated epoch) cannot be shown in this $y=0$ slice.}
\label{fig_phaseMap1}
\end{figure}

Among the prominent field theories for gravity, we will consider the scalar-tensor theories with derivative self-interactions and non-minimal couplings. 
Promoting Galileon interactions into curved space-time revealed the necessity of non-minimal couplings and the rediscovery of Horndeski interactions. They constitute the most general Lagrangian for a scalar field on curved space-time with second order equations of motion despite the presence of derivative self-interactions. The action reads $\mathcal{S}=\int d^4x \sqrt{-g}\mathcal{L}_i$ where the individual Lagrangians have to be of the following form
\begin{eqnarray}\label{eqHorndeskiLis}
\mathcal{L}_2&=&G_2(\pi,X) \\
\mathcal{L}_3&=&G_3(\pi,X)[\Pi] \nonumber\\
\mathcal{L}_4&=&G_4(\pi,X)R +G_{4,X}([\Pi]^2-[\Pi^2]) \nonumber\\
\mathcal{L}_5&=&G_5(\pi,X)G_{\mu\nu}\Pi^{\mu\nu}-\frac{G_{4,X}}{6}([\Pi]^3-3[\Pi][\Pi^2]+2[\Pi^3]) \nonumber\;,
\end{eqnarray}
where $X$ stands for the standard kinetic term $X=-\frac12\partial_\mu\pi\partial^\mu\pi$ and $\Pi_{\mu\nu}=\nabla_\mu\partial_\nu$, $[\Pi]=\Pi_\mu{}^\mu$. In the fourth and fifth Lagrangian we see the presence of non-minimal couplings to gravity through the Ricci scalar and the Einstein tensor and their relative tunings to the derivative self-interactions are essentiell to guarantee the second order nature of the equations of motion. Since the fourth order Lagrangian already captures all the particularities of derivative self-interactions and non-minimal couplings, we will only
study the Horndeski Lagrangian including cubic and quartic interactions in this work, hence our action will be
\begin{eqnarray}\label{eqHorndeski}
\mathcal{S}&=&\int d^4x \sqrt{-g}\left\{ \mathcal{L}_2+\mathcal{L}_3+\mathcal{L}_4\right\}\;.
\end{eqnarray}
Within this class of non-minimally coupled scalar-tensor theories, we need to make an Ansatz for the general coefficient functions $G_{2, 3, 4}$ in order to be able to study the presence of concrete self-accelerating models. For an exhaustive classification of possible interesting Ansatz we refer to the review article \cite{Kase:2018aps}. Our aim is to consider one of the simplest Ansatz, which gives rise to self-accelerating solution and apply the Swampland condition on it. Such a realization is for instance given by (detailed information can be found in \cite{Kase:2015zva})
\begin{align}
G_2(\pi,X)&=(1-6\alpha^2)f(\pi)X-V(\pi)\;,\\
G_3(\pi,X)&=\alpha_3 X\;,\\
G_4(\pi,X)&=\frac{\mpl}{2}f(\pi)+\alpha_4X^2\;,
\end{align}
where the function $f(\pi)$ is assumed to be $f(\pi)=e^{-2\alpha(\pi-\pi_0)/\mpl}$ and $\alpha$,  $\alpha_3$, $\alpha_4$ and $\pi_0$ are constant parameters.
The action $\mathcal{S}_{\rm m}$ of the standard matter fields has to be added. Let us now introduce the following dynamical variables: 
\begin{align}
  x_1 &= \dot\pi/(\sqrt{6}H\mpl)\;,\nonumber\\
  x_2 &= V(\pi)/(3\mpl^2H^2f(\pi))\;,\nonumber\\
  x_3 &= 6\alpha_3\dot{\pi}^3/(\mpl^2Hf(\pi))\;,\nonumber\\
  x_4 &= 10\alpha_4\dot{\pi}^4/(\mpl^2f(\pi))\;,\nonumber\\
  \Omega_r &= y^2 = \rho_r/(3\mpl^2H^2f(\pi))\;,\quad\text{and} \nonumber\\
  \lambda &= -\mpl V'/V\;.
\end{align}
The background equations of motion of the system can then be brought into the autonomous form \cite{Kase:2015zva}
\begin{align}\label{EOMquintessence_a}
\frac{dx_1}{dN}&=x_1(\epsilon_\pi-h)\;,\\
\frac{dx_2}{dN}&=x_2(\sqrt{6}(2\alpha-\lambda)x_1-2h)\;,\\
\frac{dx_3}{dN}&=x_3(2\sqrt{6}\alpha x_1+3\epsilon_\pi-h)\;,\\
\frac{dx_4}{dN}&=x_4(2\sqrt{6}\alpha x_1+4\epsilon_\pi)\;,\\
\frac{dy}{dN}&=y(\sqrt{6}\alpha x_1-2-h)\;,\\
\frac{d\lambda}{dN}&=\sqrt{6}\lambda^2 x_1\,,
\label{EOMquintessence_b}\end{align}
with the cumbersome expressions for $h=\dot{H}/H^2$ and $\epsilon_\pi=\ddot\pi/(H\dot\pi)$ given in Appendix \ref{secApp1}. Furthermore, we have
\begin{equation}
\Omega_m=1-(1-6\alpha^2)x_1^2-2\sqrt{6}\alpha x_1-x_2-x_3-\Omega_r
\end{equation}
for the matter-density parameter.  On the other hand, the dark-energy equation of state $w\equiv P/\rho$ satisfies
\begin{equation}
w=\frac{w_{\rm eff}-(\Omega_r/3)(f/f_0)}{1-(\Omega_m-\Omega_r/3)(f/f_0)}\;,
\end{equation}
with $f_0$ being the present value of $f(\pi)$ and $w_{\rm eff}$ the effective equation of state $w_{\rm eff}\equiv-1-2\dot{H}/(3H^2)$, 
\begin{equation}
  w_{\rm eff}=-1-\frac{2h}{3}\;.
\end{equation} 
Note that the presence of the cubic and quartic derivative self-interactions are encoded in the dynamical variables $x_3$ and $x_4$. In the absence of these interactions, i.e.\ for $x_3=0$ and $x_4=0$, we have a k-essence field coupled to the Ricci scalar. In this case, the autonomous system admits the critical points $(x_1, x_2, y)$ 
\begin{align}
  I:   \quad (&0,0,1)\;,\\
  II:  \quad (&(\sqrt{6}\alpha-1)^{-1},0,0)\;,\nonumber\\
  III: \quad (&(\sqrt{6}\alpha+1)^{-1},0,0)\;,\nonumber\\
  IV:  \quad (&\sqrt{2/3}\alpha(2\alpha^2-1)^{-1},0,0)\;,\nonumber\\
  V:   \quad (&2\sqrt{2/3}/\lambda,4/(3\lambda^2),\sqrt{-4+(\lambda-4\alpha)^2}/\lambda)\;,\nonumber\\
  VI:  \quad (&\sqrt{3/2}/\lambda,(3+2\alpha(\lambda-3\alpha))/(2\lambda^2),0)\;,\quad\text{and} \nonumber\\
  VII: \quad (&(\sqrt{6}(\alpha+(\lambda-4\alpha)^{-1})^{-1}, \nonumber\\
              &(-6+(\lambda-4\alpha)^2)/(6(1+\alpha(\lambda-4\alpha))^2),0)\;.\nonumber
\end{align}
The critical point $I$ corresponds to the radiation-dominated epoch, whereas the critical points $IV$ and $VII$ represent the matter- and scalar-field-dominated epochs, respectively. This is shown in Fig.\ \ref{fig_phaseMap1}.

\begin{figure}[t]
  \includegraphics[height=3.0in,width=3.0in]{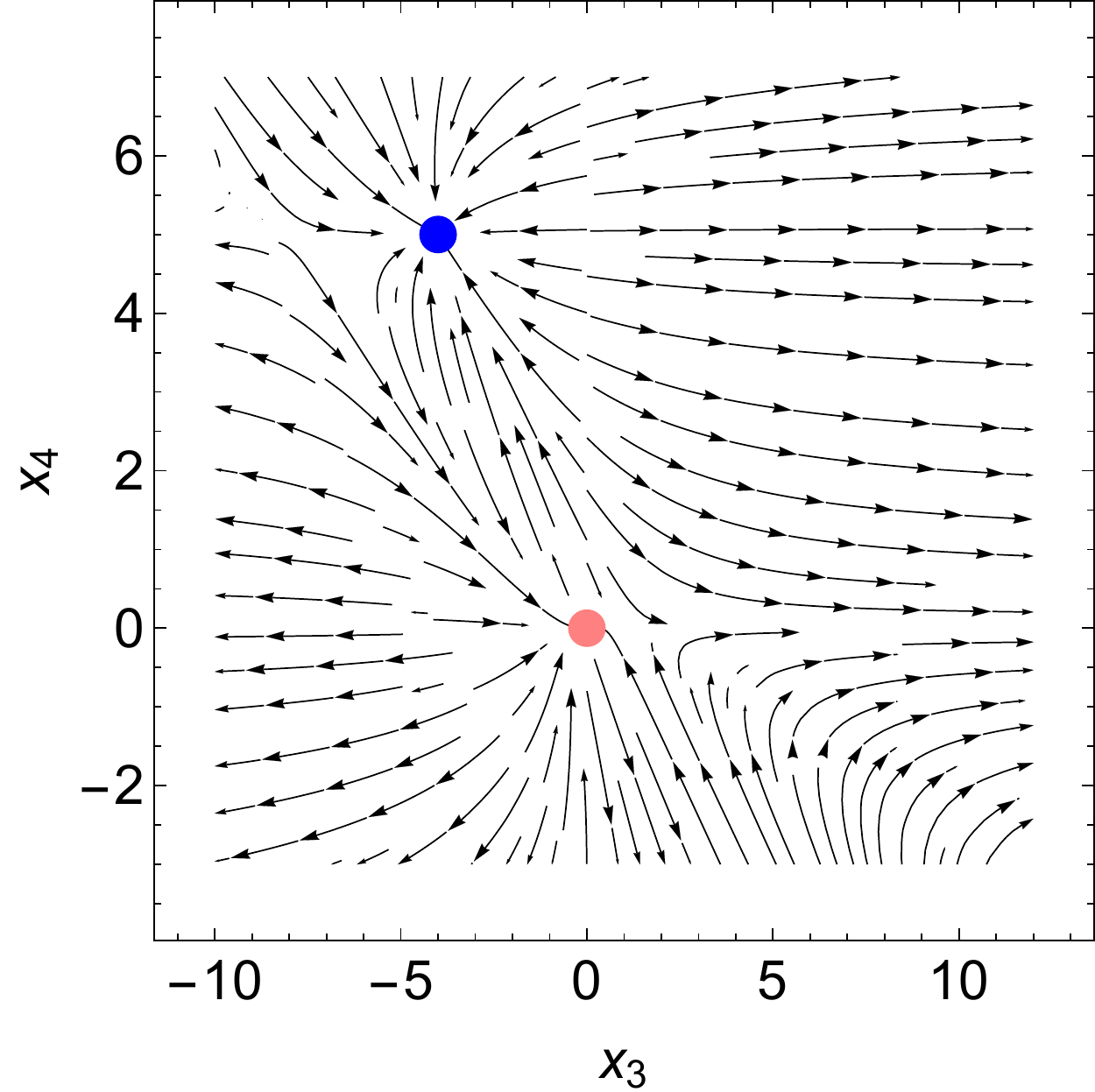}
\caption{The figure shows the phase map of the autonomous system for the model with $x_1=0$ and $x_2=0$. The trajectories of $\{ x_3,x_4\}$ on the $y=0$ plane are shown for $\alpha=-0.02$ and $\lambda=0.9$. The critical point (blue) represents the scalar-field-dominated epoch with $x_4\gg x_3\gg x_1^2$.}
\label{fig_phaseMap2}
\end{figure}

The previous model did not include the important derivative self-interactions. Additional cosmological background evolutions arise after reintroducing the parameters $\alpha_3\ne0$ and $\alpha_4\ne0$. The number of critical points increases significantly. Their exact expressions as well as the five-dimensional phase map become cumbersome to illustrate. Instead, we show a particular example of the phase map with the possible trajectories in the $x_3$ and $x_4$ plane, where these derivate self-interactions dominate, $x_4\gg x_3\gg x_1^2$. In order to show some of the critical points, we have chosen the points of intersections in the higher dimensional field space as $y=0$, $x_1=0$ and $x_2=0$. This example is shown in Fig.\ \ref{fig_phaseMap2}.

These effective field theories admit a rich phenomenology and many models for an accelerated universe relevant for the cosmological evolution. We will now use the additional constraint from the Swampland conjecture in order to further restrict the Horndeski interactions.
 An effective field theory consistent with string theory has to satisfy
\begin{itemize}  
  \item the derivative of the scalar-field potential has to satisfy the lower bound $|V'|/V>c\sim\mathcal{O}(1)$ \cite{Obied:2018sgi}; and
  \item the range traversed by a scalar field is bounded by $|\Delta\pi|<d\sim\mathcal{O}(1)$ in reduced Planck units \cite{Ooguri:2006in}.
\end{itemize}
The former conjecture has also a refined version, 
according to which if the condition on the first derivative is not satisfied, then a condition on the second derivative  ${\rm min}(V'')/V\le-\tilde{c}\sim\mathcal{O}(1)$ \cite{Ooguri:2018wrx} needs to be met. This applies in particular to models where the scalar field is close to a local maximum of the potential. Here, we are interested in evolving configurations like those used to obtain Quintessence models in which the condition on the second derivative is violated and hence the criterium on the first derivative needs to be met.
The second conjecture is automatically satisfied for the relevant cosmological evolution, since we do not require many e-foldings of accelerated expansion. We will focus on the bound $|V'|/V>c\sim\mathcal{O}(1)$ in this work.

In the following, we will solve the background equations \eqref{EOMquintessence_a} numerically, construct specific cosmological models, and compare them with cosmological observations.

\section{Observational bounds}

\subsection{Solutions for the dynamical variables}

We intend to apply the observational bounds from SNeIa, CMB, BAO and $H_0$ measurements as well as the forecast for Euclid on the Horndeski Lagrangian together with the Swampland conjecture on the potential term. For doing so, we will solve the underlying background equations numerically. These will sensitively depend on the choice of the initial conditions. They will be chosen in such a way that the resulting cosmological evolution has the right radiation-, matter-, and scalar-field dominated phases. In order to not strongly modify the distance to the last-scattering surface of the Cosmic Microwave Background, we will impose $|\alpha|<0.1$. On the other hand, the absence of fifth forces on the Solar-System scales forces us to set $|\alpha|\gtrsim0.001$ \cite{Kase:2015zva}. Similarly, the presence of a proper matter-dominated phase requires $x_4(z=0)\lesssim 10^{-6}$ and $x_3(z=0)\lesssim 10^{-4}$. The presence of the derivative self-interactions strongly modifies the dynamics of the scalar field. In order to reproduce the right cosmological evolution, the interactions have to be significantly tuned. Such a tuning is realized for the set of initial conditions satisfying $x_4\gg x_3\gg x_1^2$ (for instance $x_1=1.66\times10^{-13}$, $x_2=7.02\times10^{-25}$, $x_3=6.51\times10^{-22}$. $x_4=1.54\times10^{-18}$, $x_5=0.99985$ at redshift $z=1.02\times10^{7}$ \cite{Kase:2015zva}). We will keep these initial conditions throughout this work and scan only the two-dimensional parameter space $(\lambda,\alpha)$.

\begin{figure}[t!]
  \includegraphics[width=\hsize]{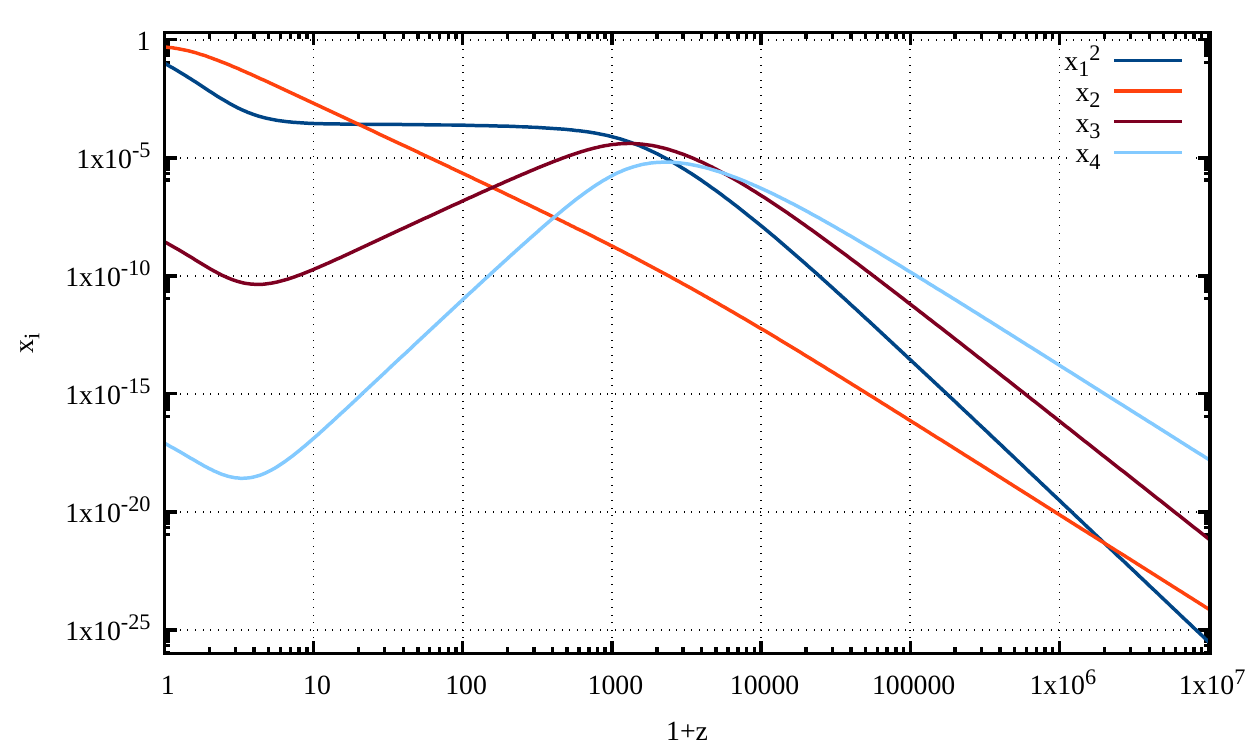}
  \includegraphics[width=\hsize]{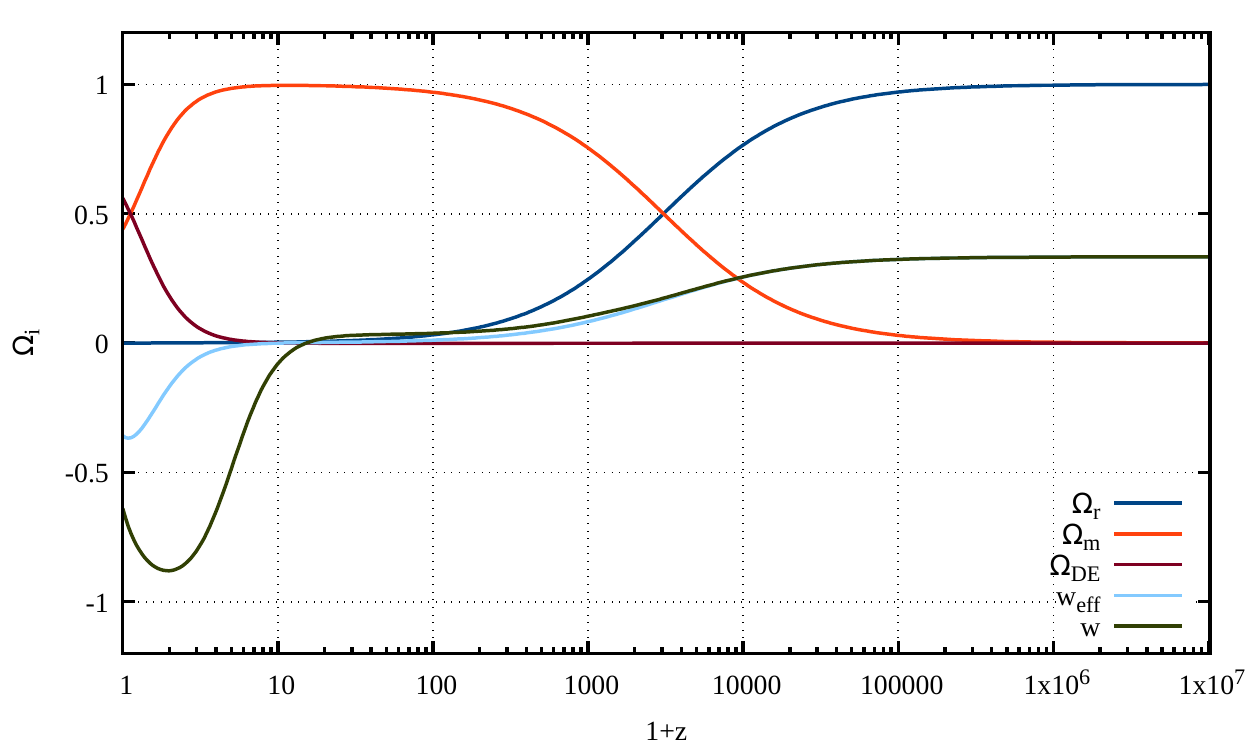}
\caption{The figure shows the numerical solutions for (top) $x_i$ and (bottom) $\Omega_i$ as a function of redshift $1+z$ for the model parameter $\alpha=-0.02$ and the initial conditions (at $z=1.02\times10^{7}$) $x_1=1.66\times10^{-13}$, $x_2=7.02\times10^{-25}$, $x_3=6.51\times10^{-22}$. $x_4=1.54\times10^{-18}$, $x_5=0.99985$ and $\lambda=0.9$. One clearly sees the transition between the radiation-, matter-, and scalar-field-dominated phases throughout the cosmic evolution.}
\label{fig_numMod}
\end{figure}

Since the de-Sitter Swampland conjecture demands $|V'|/V\ge c\sim\mathcal{O}(1)$, we will only consider the cases $\lambda(\pi)\ge c\sim\mathcal{O}(1)$. In Fig. \ref{fig_numMod} we show an example of our numerical solutions for the dynamical variables of the autonomous system \eqref{EOMquintessence_a} for $|\alpha|=0.02$ and $\lambda=0.9$. It can be clearly seen that the evolution starts from the initial conditions satisfying $x_4\gg x_3\gg x_1^2$. During the radiation-dominated phase, $x_1^2$ grows faster than $x_4$ and $x_3$ and outpaces them by the end of the radiation-dominated epoch. Then, the matter-dominated phase takes over and $x_2$ starts dominating. During this period, we have $x_2>x_1^2> x_3> x_4$ and the derivative self-interactions $x_3$ and $x_4$ decrease significantly. Next, the $\pi$-dominated phase starts. During this epoch, the cosmic acceleration starts only once $x_2$ also overtakes $\Omega_m$. While the universe undergoes the phase of acceleration with almost constant $H$, the ratios between the derivative self-interactions and $x_1$ are kept nearly constant. The evolution of the density parameters $\Omega_{\rm DE}$, $\Omega_{\rm m}$ and $\Omega_{\rm r}$ together with the equation of state parameter $w$ of the dark energy are also shown in the figure. It illustrates that the cosmic evolution follows the radiation-, matter-, and $\pi$-dominated phases throughout the history of the universe.

\subsection{Observational upper bound on $w(z)$}

The observational constraints can be approximately represented by a confidence ellipse with semi-axes $a$ and $b$, described by $x^2/a^2+y^2/b^2=1$ in its principal-axis frame with the origin of $\vec x = (x,y)^\top$ on the centre of the ellipse, with the coordinates represented by $x$ and $y$. If the coordinate frame is rotated by an orthogonal matrix $R$ and has its origin shifted to the point $\vec x_\mathrm{c} = (x_\mathrm{c}, y_\mathrm{c})^\top$, the ellipse is described by $(\vec x-\vec x_\mathrm{c})^\top M(\vec x-\vec x_\mathrm{c}) = 1$ with $M = R\,\mathrm{diag}(a^{-2},b^{-2})R^\top$.

In the CPL parameterization \cite{Chevallier:2000qy} of $w$ as a function of redshift $z$,
\begin{equation}
  w(z) = w_0+\frac{z}{1+z}\,w_\mathrm{a} =: w_0+\zeta w_\mathrm{a}\;,
\label{CPL}
\end{equation}
we approximate the measured constraints on $w_0$ and $w_\mathrm{a}$ by elliptical uncertainty contours in the $(w_0, w_\mathrm{a})$ plane around the best-fitting point $\vec w_\mathrm{c} = (w_0, w_\mathrm{a})_\mathrm{c}^\top$, modelled by the inverse Fisher matrix $M$ with its independent elements $M_{11}$, $M_{12} = M_{21}$ and $M_{22}$. These confidence contours then satisfy the equation
\begin{equation}
  (\vec w-\vec w_\mathrm{c})^\top M(\vec w-\vec w_\mathrm{c}) = 1\;.
\label{ellipse}
\end{equation}
If we parameterize the contour by an angle $\phi\in[0,2\pi)$ and write it in the form
\begin{equation}
  \vec w(\phi) = \vec w_\mathrm{c} + \delta w(\cos\phi,\sin\phi)^\top\;,
\end{equation} 
Eq.\ (\ref{ellipse}) implies $\delta w = q^{-1/2}$ with
\begin{equation}
  q = M_{11}\cos^2\phi+2M_{12}\cos\phi\sin\phi+M_{22}\sin^2\phi\;.
\end{equation}
Inserting
\begin{align}
  w_0(\phi) &= w_\mathrm{0c}+\delta w\cos\phi\;,\nonumber\\
  w_\mathrm{a}(\phi) &= w_\mathrm{0a}+\delta w\sin\phi
\end{align}
into the CPL parameterization (\ref{CPL}) gives
\begin{equation}
  w(z,\phi) = a(\phi)+\zeta \frac{b(\phi)}{\sqrt{q}}
\label{w_phi}
\end{equation}
with $a(\phi) = w_\mathrm{0c}+\zeta w_\mathrm{ac}$ and $b(\phi) = \cos\phi+\zeta\sin\phi$. We can then find the upper limit on $w(z)$ allowed by the observational constraints analytically by searching for $\max_{\phi} w(z,\phi)$. Taking the derivative of $w(z,\phi)$ with respect to $\phi$ and equating the result to zero gives $2qb'-bq' = 0$, where the prime denotes the derivative with respect to $\phi$. This is a third-order polynomial in $\tan\phi$ with the only real solution
\begin{equation}
  \tan\phi =: \tau = \frac{M_{12}-\zeta M_{11}}{\zeta M_{12}-M_{22}}
\end{equation} 
for $\tan\phi$. Since the tangent is $\pi$-periodic, this solution contains both maxima and minima of $w(z,\phi)$, with the maximum identified by the solution with positive $\cos\phi$. Inserting this solution into (\ref{w_phi}) gives the upper bound
\begin{equation}
  w_\mathrm{max}(z) = w_\mathrm{0c}+\zeta w_\mathrm{ac}+
  \frac{|1+\zeta\tau|}{\sqrt{M_{11}+2M_{12}\tau+M_{22}\tau^2}}
\label{w_max}
\end{equation}
on $w(z)$, characterized by the elements of the inverse Fisher matrix $M$. We will apply this analytical expression to the corresponding 1- and 2-$\sigma$ ellipses enclosing the domain in the $(w_0, w_\mathrm{a})$ plane allowed by the observations.

\subsection{Constraints in the $(\lambda, \alpha)$ plane}

\begin{figure}[h!]
  \includegraphics[width=\hsize]{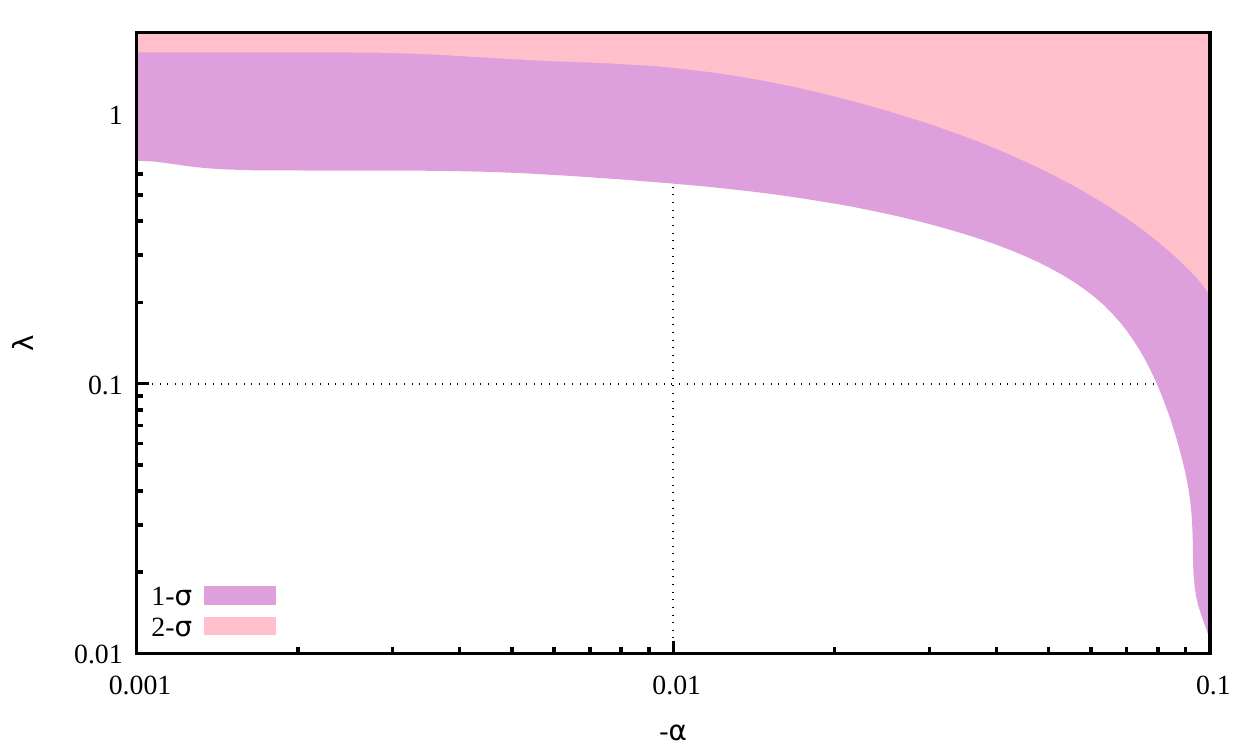}
\caption{The figure shows the allowed parameter space $(\lambda,\alpha)$ after comparing the upper bound on the reconstructed equation of state $w(z)$ of the 1- and 2-$\sigma$ contours of Fig.\ 21 of \cite{Scolnic:2017caz} with our numerical solutions of the Horndeski model. Luminal propagation speed for gravitational waves is assumed, i.e. we set $\alpha_4=0$.}
\label{fig_current}
\end{figure}

In order to compare our numerical solutions for the equation of state parameter $w$ for different values of $(\lambda,\alpha)$ with the observational constraints obtained in \cite{Scolnic:2017caz}, we estimate the inverse Fisher matrix $M$ from the 1- and 2-$\sigma$ contours of Fig.\ 21 in \cite{Scolnic:2017caz}. In this way, we can directly compare the empirical upper bound on the equation of state parameter $w$ as a function of redshift from (\ref{w_max}) with our numerical solutions of the background equations. The result is illustrated in Fig. \ref{fig_current}. The comparison of the observational uncertainties on $w_0$ and $w_\mathrm{a}$ with the regime still allowed by the string Swampland criteria requires $\lambda\lesssim 0.6$ for a wide range of values for $\alpha$.

\begin{figure}[h!]
  \includegraphics[width=\hsize]{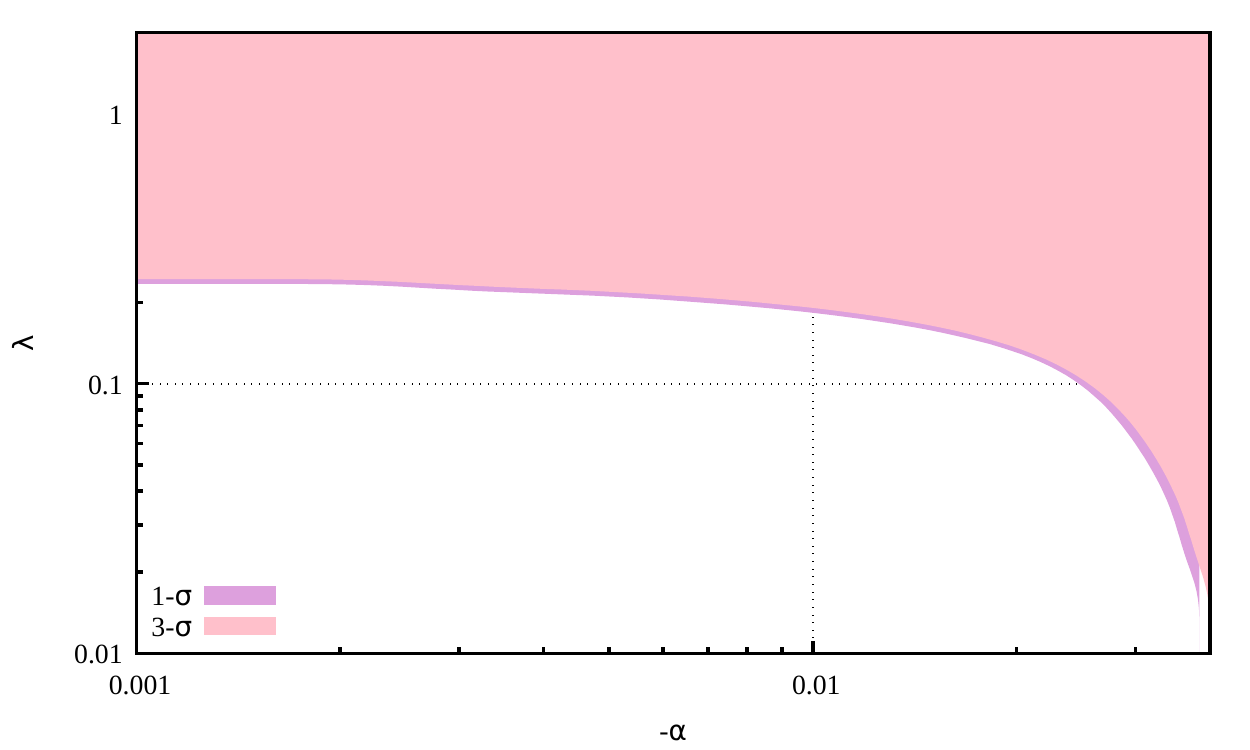}
\caption{The figure illustrates the allowed parameter space $(\lambda,\alpha)$ of Euclid type stage-4 experiment.}
\label{fig_stage4}
\end{figure}

Similarly, we can use the near-future limits of Stage-4 surveys to obtain tighter constraints on the allowed Horndeski models within the Swampland criteria. The outcome is shown in Fig. \ref{fig_stage4}, where the prospective 1- and 3-$\sigma$ upper bounds on $w_0$ and $w_\mathrm{a}$ were taken from the Euclid Definition Study Report \cite{stage4}, and the orientation of the inverse Fisher matrix was assumed to be the same as in the current observational constraints. As we can see, the planned Stage-4 surveys exemplified by Euclid can already be expected to lower the allowed values for $\lambda$ to $\lambda\lesssim 0.2$. With this, the entire class of Horndeski dark-energy models would be pushed into an uncomfortable corner.

The significant discovery of two merging neutron stars has significantly constrained the propagation speed $c_T$ of gravitational waves to be very close to the speed of light $|c_T/c-1|<10^{-15}$. Horndeski theories including the quartic and quintic interactions are tightly constrained by this observation. Specifically, the non-minimal couplings to the gravity sector in $\mathcal{L}_4$ and $\mathcal{L}_5$ (which we did not consider in \eqref{eqHorndeski}) contribute to the propagation speed of the tensor modes as
\begin{equation}
c_T^2=\frac{2G_4-2XG_{5,\pi}-2XG_{5,X}\ddot{\pi}}{4q_T}\,,
\end{equation}
where $q_T$ represents the kinetic term of the tensor modes
\begin{equation}
q_T=\frac14(2(G_4-2XG_{4,X})-2X(G_{5,X}\ddot\pi))\,.
\end{equation}
Hence, luminal propagation of gravitational waves in Horndeski models would require $G_{4,X}=0$ and $G_5=0$ (see for instance \cite{Ezquiaga:2017ekz} and some reviews \cite{Heisenberg:2018vsk,Kase:2018aps,JMEandMZ}). We have taken this into account when we produced the allowed parameter space in figures \ref{fig_current} and \ref{fig_stage4}.

\section{Conclusion}

In this work, we have applied the de-Sitter Swampland conjecture to Horndeski scalar-tensor theories, which represent a prominent class of alternative theories of gravity based on an additional scalar field. The defining properties of the Horndeski interactions are that they contain derivative self-interactions and non-minimal couplings, but still give rise to second-order equations of motion. The Quintessence model corresponds to just a restricted sub-class of this general scalar-tensor theories. The presence of these derivative self-interactions crucially influences the dynamics of the scalar field. The requirement of the appropriate cosmological evolution strengthens the implications of the de-Sitter Swampland conjecture. The distinctive interactions arise in this cubic and quartic Horndeski Lagrangians, which we encoded in $x_3$ and $x_4$. The dynamical background equations rely strongly on the choice of the initial conditions. In order for these higher-order interactions not to be too small and to have appreciable effects with $x_4\gg x_3\gg x_1^2$, the initial conditions have to be significantly tuned for the appropriate cosmology. We have chosen such conditions throughout this work, such that the successive epochs of radiation-, matter- and $\pi$-domination were ensured. Conversely, this leaves little room for the slope of the potential. Hence, the de-Sitter Swampland conjecture gives rise to tighter constraints within the Horndeski dark energy models.

\section*{Acknowledgements}
We are grateful for useful discussions with Cumrun Vafa. LH is supported by funding from the European Research Council (ERC) under the European Union’s Horizon 2020 research and innovation programme grant agreement No 801781 and by the Swiss National Science Foundation grant 179740. RB is supported in part by funds from NSERC and from the Canada Research Chair program. 

\appendix

\section{Variables of the dynamical system}\label{secApp1}

\begin{widetext}
\begin{eqnarray}
\epsilon_{\phi}
&=&
[3\sqrt{6}x_1^2 \{ 5x_3-20+2q^2 (20+5x_3-4x_4)+12x_4 \}
+12q x_1^3 \{ 25-2x_4+6q^2(2x_4-5 )\}+\sqrt{6} \{ -24x_2x_4\nonumber \\
& &+3x_3 (x_4-5x_2-5)
+(5x_3+8x_4)x_5^2 \} 
-12x_1 \{ \lambda x_2(x_4-5)+5q(1+3x_2+2x_3
+3x_4-x_5^2)\}]/(\sqrt{6}{\cal D}),\label{ddotphi}\\
h
&=& -[30(1-8q^2+12q^4)x_1^4+15x_3^2+2\sqrt{6}q
(6q^2-1)x_1^3 (5x_3+8x_4-10) 
+\sqrt{6}x_1 \{ 5x_3(2q-\lambda x_2)
+8x_4(q-\lambda x_2) \}\nonumber \\
& &+10x_3 (3-3x_2+3x_4+x_5^2)
+12x_4(3-3x_2+x_4+x_5^2)
+2x_1^2 (15-15x_2+30x_3+39x_4+5x_5^2) \nonumber \\
& &-60q x_1^2 \{ \lambda x_2-q(1+3x_2-2x_3-3x_4-x_5^2) \}]
/{\cal D}\,,\label{dotH} \\
{\cal D}
&=&
5x_3(4+x_3)+12(2+x_3)x_4+8x_4^2+4\sqrt{6}qx_1 
(5x_3+8x_4)+4x_1^2 \{5+(6q^2-1)x_4 \}\,.
\end{eqnarray}
\end{widetext}

%%%%%%%%%%%%%%%%


\begin{thebibliography}{99}

%\cite{Horndeski:1974wa}
\bibitem{Horndeski:1974wa} 
  G.~W.~Horndeski,
  %``Second-order scalar-tensor field equations in a four-dimensional space,''
  Int.\ J.\ Theor.\ Phys.\  {\bf 10}, 363 (1974).
  C.~Deffayet, G.~Esposito-Farese and A.~Vikman,
  %``Covariant Galileon,''
  Phys.\ Rev.\ D {\bf 79}, 084003 (2009)
 % doi:10.1103/PhysRevD.79.084003
  [arXiv:0901.1314 [hep-th]].
  C.~Deffayet, S.~Deser and G.~Esposito-Farese,
  %``Generalized Galileons: All scalar models whose curved background extensions maintain second-order field equations and stress-tensors,''
  Phys.\ Rev.\ D {\bf 80} (2009) 064015
  %%doi:10.1103/PhysRevD.80.064015
  [arXiv:0906.1967 [gr-qc]].
  C.~Deffayet, O.~Pujolas, I.~Sawicki and A.~Vikman,
  %``Imperfect Dark Energy from Kinetic Gravity Braiding,''
  JCAP {\bf 1010}, 026 (2010)
%  doi:10.1088/1475-7516/2010/10/026
  [arXiv:1008.0048 [hep-th]].
  C.~Deffayet, X.~Gao, D.~A.~Steer and G.~Zahariade,
  %``From k-essence to generalised Galileons,''
  Phys.\ Rev.\ D {\bf 84}, 064039 (2011)
  %%doi:10.1103/PhysRevD.84.064039
  [arXiv:1103.3260 [hep-th]];
    C.~de Rham and L.~Heisenberg,
  %``Cosmology of the Galileon from Massive Gravity,''
  Phys.\ Rev.\ D {\bf 84}, 043503 (2011)
 % doi:10.1103/PhysRevD.84.043503
  [arXiv:1106.3312 [hep-th]];
  T.~Kobayashi, M.~Yamaguchi and J.~Yokoyama,
  %``Generalized G-inflation: Inflation with the most general second-order field equations,''
  Prog.\ Theor.\ Phys.\  {\bf 126}, 511 (2011)
   [arXiv:1105.5723 [hep-th]].



%\cite{Heisenberg:2018vsk}
\bibitem{Heisenberg:2018vsk} 
  L.~Heisenberg,
  ``A systematic approach to generalisations of General Relativity and their cosmological implications,''
  arXiv:1807.01725 [gr-qc].

\bibitem{OoguriVafa}
H.~Ooguri and C.~Vafa,
  ``On the Geometry of the String Landscape and the Swampland,''
  Nucl.\ Phys.\ B {\bf 766}, 21 (2007)
  %doi:10.1016/j.nuclphysb.2006.10.033
  [hep-th/0605264].
 
 \bibitem{Obied:2018sgi} 
  G.~Obied, H.~Ooguri, L.~Spodyneiko and C.~Vafa,
  %``De Sitter Space and the Swampland,''
  arXiv:1806.08362 [hep-th]. 
  
  \bibitem{noLambda}
  U.~H.~Danielsson and T.~Van Riet,
  %``What if string theory has no de Sitter vacua?,''
  arXiv:1804.01120 [hep-th];
G.~Dvali and C.~Gomez,
 % ``Quantum Exclusion of Positive Cosmological Constant?,''
  Annalen Phys.\  {\bf 528}, 68 (2016)
%  doi:10.1002/andp.201500216
  [arXiv:1412.8077 [hep-th]];
A.~Castro, N.~Lashkari and A.~Maloney,
 % ``A de Sitter Farey Tail,''
  Phys.\ Rev.\ D {\bf 83}, 124027 (2011)
 % doi:10.1103/PhysRevD.83.124027
  [arXiv:1103.4620 [hep-th]];
 S.~Sethi,
%  ``Supersymmetry Breaking by Fluxes,''
  arXiv:1709.03554 [hep-th];
  J.~McOrist and S.~Sethi,
%  ``M-theory and Type IIA Flux Compactifications,''
  JHEP {\bf 1212}, 122 (2012)
%  doi:10.1007/JHEP12(2012)122
  [arXiv:1208.0261 [hep-th]];
J.~J.~Heckman, C.~Lawrie, L.~Lin and G.~Zoccarato,
  %``F-theory and Dark Energy,''
  arXiv:1811.01959 [hep-th];
  J.~J.~Heckman, C.~Lawrie, L.~Lin, J.~Sakstein and G.~Zoccarato,
  %``Pixelated Dark Energy,''
  arXiv:1901.10489 [hep-th].

  
    
%\cite{CosmoImpl}
\bibitem{CosmoImpl} 
  P.~Agrawal, G.~Obied, P.~J.~Steinhardt and C.~Vafa,
 % ``On the Cosmological Implications of the String Swampland,''
  arXiv:1806.09718 [hep-th];
L.~Heisenberg, M.~Bartelmann, R.~Brandenberger and A.~Refregier,
  %``Dark Energy in the Swampland,''
  arXiv:1808.02877 [astro-ph.CO];
  W.~H.~Kinney, S.~Vagnozzi and L.~Visinelli,
  %``The Zoo Plot Meets the Swampland: Mutual (In)Consistency of Single-Field Inflation, String Conjectures, and Cosmological Data,''
  arXiv:1808.06424 [astro-ph.CO];
  Y.~Akrami, R.~Kallosh, A.~Linde and V.~Vardanyan,
  %``The landscape, the swampland and the era of precision cosmology,''
  arXiv:1808.09440 [hep-th].
  L.~Heisenberg, M.~Bartelmann, R.~Brandenberger and A.~Refregier,
  %``Dark Energy in the Swampland II,''
  arXiv:1809.00154 [astro-ph.CO];
  D.~Wang,
  %``The multi-feature universe: large parameter space cosmology and the swampland,''
  arXiv:1809.04854 [astro-ph.CO];
  H.~Fukuda, R.~Saito, S.~Shirai and M.~Yamazaki,
  %``Phenomenological Consequences of the Refined Swampland Conjecture,''
  arXiv:1810.06532 [hep-th].
 

\bibitem{recent}
F.~Denef, A.~Hebecker and T.~Wrase,
 % ``The dS swampland conjecture and the Higgs potential,''
  arXiv:1807.06581 [hep-th];
  D.~Andriot,
%  ``On the de Sitter swampland criterion,
  arXiv:1806.10999 [hep-th];
  C.~Roupec and T.~Wrase,
 % ``de Sitter extrema and the swampland,''
  arXiv:1807.09538 [hep-th];
  %%CITATION = ARXIV:1807.09538;%%
  A.~Kehagias and A.~Riotto,
%  ``A note on Inflation and the Swampland,''
  arXiv:1807.05445 [hep-th];
  %%CITATION = ARXIV:1807.05445;%%
  J.~L.~Lehners,
%  ``Small-Field and Scale-Free: Inflation and Ekpyrosis at their Extremes,''
  arXiv:1807.05240 [hep-th];
C.~Han, S.~Pi and M.~Sasaki,
  %``Quintessence Saves Higgs Instability,''
  arXiv:1809.05507 [hep-ph];
  C.~M.~Lin, K.~W.~Ng and K.~Cheung,
  %``Chaotic inflation on the brane and the Swampland Criteria,''
  arXiv:1810.01644 [hep-ph].


  

  
  %\cite{Nicolis:2008in}
\bibitem{Nicolis:2008in} 
  A.~Nicolis, R.~Rattazzi and E.~Trincherini,
  %``The Galileon as a local modification of gravity,''
  Phys.\ Rev.\ D {\bf 79}, 064036 (2009)
%  doi:10.1103/PhysRevD.79.064036
  [arXiv:0811.2197 [hep-th]].
  
  %\cite{deRham:2010kj}
\bibitem{deRham:2010kj} 
  C.~de Rham, G.~Gabadadze and A.~J.~Tolley,
  %``Resummation of Massive Gravity,''
  Phys.\ Rev.\ Lett.\  {\bf 106}, 231101 (2011)
  doi:10.1103/PhysRevLett.106.231101
  [arXiv:1011.1232 [hep-th]].
  %%CITATION = doi:10.1103/PhysRevLett.106.231101;%%
  %1059 citations counted in INSPIRE as of 16 Oct 2018

\bibitem{HeisenbergProca} 
L.~Heisenberg,
%``Generalization of the Proca Action,''
JCAP {\bf 1405}, 015 (2014)
%doi:10.1088/1475-7516/2014/05/015
[arXiv:1402.7026 [hep-th]];
E.~Allys, P.~Peter and Y.~Rodriguez,
%``Generalized Proca action for an Abelian vector field,''
JCAP {\bf 1602}, 004 (2016).
%doi:10.1088/1475-7516/2016/02/004
[arXiv:1511.03101 [hep-th]];
J.~Beltran~Jimenez and L.~Heisenberg,
%``Derivative self-interactions for a massive vector field,''
Phys.\ Lett.\ B {\bf 757}, 405 (2016).
%doi:10.1016/j.physletb.2016.04.017
[arXiv:1602.03410 [hep-th]];
  G.~Tasinato,
  %``Cosmic Acceleration from Abelian Symmetry Breaking,''
  JHEP {\bf 1404}, 067 (2014)
[arXiv:1402.6450 [hep-th]];
  L.~Heisenberg, R.~Kase and S.~Tsujikawa,
  %``Beyond generalized Proca theories,''
  Phys.\ Lett.\ B {\bf 760}, 617 (2016)
  [arXiv:1605.05565 [hep-th]].


%\cite{deRham:2011by}
\bibitem{deRham:2011by} 
  C.~de Rham and L.~Heisenberg,
  %``Cosmology of the Galileon from Massive Gravity,''
  Phys.\ Rev.\ D {\bf 84}, 043503 (2011)
 % doi:10.1103/PhysRevD.84.043503
  [arXiv:1106.3312 [hep-th]].
  
  %\cite{deRham:2017imi}
\bibitem{deRham:2017imi} 
  C.~de Rham, S.~Melville, A.~J.~Tolley and S.~Y.~Zhou,
  %``Massive Galileon Positivity Bounds,''
  JHEP {\bf 1709}, 072 (2017)
 % doi:10.1007/JHEP09(2017)072
  [arXiv:1702.08577 [hep-th]].
  
  
  %\cite{Pirtskhalava:2015nla}
\bibitem{Pirtskhalava:2015nla} 
  D.~Pirtskhalava, L.~Santoni, E.~Trincherini and F.~Vernizzi,
  %``Weakly Broken Galileon Symmetry,''
  JCAP {\bf 1509}, no. 09, 007 (2015)
 % doi:10.1088/1475-7516/2015/09/007
  [arXiv:1505.00007 [hep-th]].
  
    %\cite{Kase:2018aps}
\bibitem{Kase:2018aps}
  R.~Kase and S.~Tsujikawa,
  %``Dark energy in Horndeski theories after GW170817: A review,''
  arXiv:1809.08735 [gr-qc].
  
%\cite{Kase:2015zva}
\bibitem{Kase:2015zva} 
  R.~Kase, S.~Tsujikawa and A.~De Felice,
  %``Cosmology with a successful Vainshtein screening in theories beyond Horndeski,''
  Phys.\ Rev.\ D {\bf 93}, no. 2, 024007 (2016)
  doi:10.1103/PhysRevD.93.024007
  [arXiv:1510.06853 [gr-qc]].
  %%CITATION = doi:10.1103/PhysRevD.93.024007;%%
  %11 citations counted in INSPIRE as of 08 Oct 2018

%\cite{Ooguri:2006in}
\bibitem{Ooguri:2006in} 
  H.~Ooguri and C.~Vafa,
  %``On the Geometry of the String Landscape and the Swampland,''
  Nucl.\ Phys.\ B {\bf 766}, 21 (2007)
  %doi:10.1016/j.nuclphysb.2006.10.033
  [hep-th/0605264].


%\cite{Ooguri:2018wrx}
\bibitem{Ooguri:2018wrx} 
  H.~Ooguri, E.~Palti, G.~Shiu and C.~Vafa,
  %``Distance and de Sitter Conjectures on the Swampland,''
  arXiv:1810.05506 [hep-th];
   S.~K.~Garg and C.~Krishnan,
  %``Bounds on Slow Roll and the de Sitter Swampland,''
  arXiv:1807.05193 [hep-th].



  
  %\cite{Scolnic:2017caz}
\bibitem{Scolnic:2017caz} 
  D.~M.~Scolnic {\it et al.},
  ``The Complete Light-curve Sample of Spectroscopically Confirmed SNe Ia from Pan-STARRS1 and Cosmological Constraints from the Combined Pantheon Sample,''
  Astrophys.\ J.\  {\bf 859}, no. 2, 101 (2018)
 % doi:10.3847/1538-4357/aab9bb
  [arXiv:1710.00845 [astro-ph.CO]].
  
  %\cite{Chevallier:2000qy}
\bibitem{Chevallier:2000qy} 
  M.~Chevallier and D.~Polarski,
  ``Accelerating universes with scaling dark matter,''
  Int.\ J.\ Mod.\ Phys.\ D {\bf 10}, 213 (2001)
%  doi:10.1142/S0218271801000822
  [gr-qc/0009008];
   E.~V.~Linder,
  ``Exploring the expansion history of the universe,''
  Phys.\ Rev.\ Lett.\  {\bf 90}, 091301 (2003)
%  doi:10.1103/PhysRevLett.90.091301
  [astro-ph/0208512].
  

\bibitem{stage4} 
  R.~Laureijs {\it et al.} [EUCLID Collaboration],
  ``Euclid Definition Study Report,''
  arXiv:1110.3193 [astro-ph.CO];
 L.~Amendola {\it et al.} [Euclid Theory Working Group],
  ``Cosmology and fundamental physics with the Euclid satellite,''
  Living Rev.\ Rel.\  {\bf 16}, 6 (2013)
%  doi:10.12942/lrr-2013-6
  [arXiv:1206.1225 [astro-ph.CO]];
  P.~A.~Abell {\it et al.} [LSST Science and LSST Project Collaborations],
  ``LSST Science Book, Version 2.0,''
  arXiv:0912.0201 [astro-ph.IM];
  M.~Levi {\it et al.} [DESI Collaboration],
  ``The DESI Experiment, a whitepaper for Snowmass 2013,''
  arXiv:1308.0847 [astro-ph.CO].

\bibitem{euclidRedBook}
  Laureijs, R. et al.,
  ESA/SRE (2011) 12
  [arXiv:1110.3193]
  

\bibitem{Ezquiaga:2017ekz} 
  J.~M.~Ezquiaga and M.~Zumalacárregui,
  %``Dark Energy After GW170817: Dead Ends and the Road Ahead,''
  Phys.\ Rev.\ Lett.\  {\bf 119}, no. 25, 251304 (2017)
%  doi:10.1103/PhysRevLett.119.251304
  [arXiv:1710.05901 [astro-ph.CO]];
  %%CITATION = doi:10.1103/PhysRevLett.119.251304;%%
  %255 citations counted in INSPIRE as of 11 Apr 2019
  J.~Sakstein and B.~Jain,
  %``Implications of the Neutron Star Merger GW170817 for Cosmological Scalar-Tensor Theories,''
  Phys.\ Rev.\ Lett.\  {\bf 119}, no. 25, 251303 (2017)
  %doi:10.1103/PhysRevLett.119.251303
  [arXiv:1710.05893 [astro-ph.CO]];
  T.~Baker, E.~Bellini, P.~G.~Ferreira, M.~Lagos, J.~Noller and I.~Sawicki,
  %``Strong constraints on cosmological gravity from GW170817 and GRB 170817A,''
  Phys.\ Rev.\ Lett.\  {\bf 119}, no. 25, 251301 (2017)
 % doi:10.1103/PhysRevLett.119.251301
  [arXiv:1710.06394 [astro-ph.CO]];
  P.~Creminelli and F.~Vernizzi,
  %``Dark Energy after GW170817 and GRB170817A,''
  Phys.\ Rev.\ Lett.\  {\bf 119}, no. 25, 251302 (2017)
  %doi:10.1103/PhysRevLett.119.251302
  [arXiv:1710.05877 [astro-ph.CO]].
  

  \bibitem{JMEandMZ}
  J.~M.~Ezquiaga and M.~Zumalacárregui,
  %``Dark Energy in light of Multi-Messenger Gravitational-Wave astronomy,''
  Front.\ Astron.\ Space Sci.\  {\bf 5}, 44 (2018)
 % doi:10.3389/fspas.2018.00044
  [arXiv:1807.09241 [astro-ph.CO]].


\end{thebibliography}
\end{document}